\documentstyle[12pt]{article}
\begin{document}
\begin{titlepage}
\mbox{ }
\rightline{UCT-TP-215/94}
\rightline{June 1994}
\vspace{1cm}
\begin{center}
\begin{Large}
{\bf  Electromagnetic pion form factor at finite temperature}

\end{Large}

\vspace{1.5cm}

{\large {\bf C. A. Dominguez}}\footnote{John Simon Guggenheim
Fellow 1994-1995}

Institute of Theoretical Physics and Astrophysics, University of
Cape Town, Rondebosch 7700, South Africa\\

\vspace{.5cm}

{\large {\bf M. Loewe}}\\[0.5cm]
Facultad de Fisica, Pontificia Universidad Cat\'{o}lica
de Chile, Santiago, Chile\\

and \\

\vspace{.5cm}

{\large{\bf J. S. Rozowsky}}

Department of Physics, University of
Cape Town, Rondebosch 7700, South Africa\\
\end{center}
\vspace{.5cm}

\begin{abstract}
\noindent
The electromagnetic form factor of the pion in the space-like region, and
at finite temperature, $F_{\pi}(Q^{2},T)$, is obtained
from a QCD Finite Energy Sum Rule. The form factor decreases with
increasing T, and vanishes at some critical temperature, where the pion
radius diverges. This divergence may be interpreted as a signal for
quark deconfinement.
\end{abstract}
\end{titlepage}

\setlength{\baselineskip}{1\baselineskip}
\noindent
The possibility that QCD exhibits chiral-symmetry restoration and
quark-gluon deconfinement at finite temperature has triggered interest in
the thermal behaviour of QCD in general, and of hadronic propagators
in particular. In this respect, some time ago
a proposal was made to consider the imaginary part of any
hadronic propagator as a phenomenological order  parameter for
the deconfinement phase transition \cite{op}.
According
to this proposal, with increasing T one should
expect resonances to become broader, and stable-hadron propagators
to develop a non-zero imaginary part. This behaviour has been confirmed
later for pions, nucleons, and rho-mesons using a variety of theoretical
approaches \cite{gt}. Independent phenomenological evidence for the
deconfinement phase transition in QCD may be obtained e.g. by studying the
thermal behaviour of the electromagnetic form factor of the pion, $F_{\pi}$.
In this case one would expect the size of the pion to increase with
increasing temperature. At the critical temperature the pion radius should
presumably diverge, indicating quark-gluon deconfinement.
A recent calculation in the framework of
the Nambu-Jona Lasinio \cite{njl} model supports this scenario.\\

In this note we determine the temperature dependence of $F_{\pi}$ in
the space-like region using a Finite Energy QCD Sum Rule (FESR). The
pion form factor at $T=0$ has been extensively studied in the past with
FESR, as well as with Laplace transform QCD sum rules \cite{fpi0}.
In order to
establish some notation, as well as the $T=0$ normalization, we briefly
describe the method at $T=0$ before introducing thermal corrections.\\

The appropriate object to study is the three-point function
\begin{equation}
\Pi_{\mu \nu \lambda}(p,p',q) = i^{2} \int \; d^{4}x \; d^{4}y \;
e^{i(p'x - qy)}
< 0 |T (A_{\nu}^{\dag}(x) \; V_{\lambda} (y) \; A_{\mu}(0))|0> \,,
\end{equation}

where $A_{\mu}(x) = \bar{u}(x)\gamma_{\mu}\gamma_{5}d(x)$ is the axial-vector
current, $V_{\lambda}$ is the electromagnetic current, and $q = p'-p$ the
momentum transfer. On general analyticity grounds, the three-point function
(1) satisfies the double dispersion relation
\begin{equation}
\Pi_{\mu \nu \lambda}(p^{2},p'^{2},Q^{2}) = \frac{1}{\pi^{2}} \;
\int_{0}^{\infty} \; ds \; \int_{0}^{\infty} \; ds'
\; \frac{\rho_{\mu \nu \lambda} (s, s', Q^{2})}
	  {(s + p^{2}) (s' + p'^{2})} \,,
\end{equation}

defined up to subtractions, which are disposed of by Laplace improving
the Hilbert transform, or by considering FESR. The QCD sum rule
program at $T=0$ \cite{svz} may be phrased as follows.
The left hand side of (2) in the Euclidean region
can be calculated in QCD through the Operator Product Expansion (OPE),
to any desired order in perturbation theory, and parametrizing
non-perturbative effects in terms of vacuum expectation values of the
same quark and gluon fields entering the QCD Lagrangian. Contact with
the hadronic world is made by writing the spectral function appearing in
the right hand side of (2) in terms of all possible hadronic intermediate
states contributing to the three-point function. Normally,
a reasonable saturation of the dispersion relation is achieved
by the ground state, followed after some  threshold energy $s_{0}$
by a hadronic continuum modelled by perturbative QCD.\\

The correlator (1) involves quite a few structure functions,
associated with all the Lorentz structures that can be formed with
the available four-momenta. In principle, it should not matter which
particular structure one chooses to project the pion form factor. We
follow \cite{fpi0} in choosing the combination $P_{\mu} P_{\nu} P_{\lambda}$,
where $P = p + p'$. In this case, the hadronic spectral function
in the chiral-limit reads
\begin{equation}
\rho (s, s', Q^{2})|_{HAD} = \frac{1}{2} \; f_{\pi}^{2} F_{\pi}
(Q^{2}) \delta (s) \delta(s') +
\rho (s, s', Q^{2})|_{QCD} [1 - \theta (s_{0} - s - s')] \,,
\end{equation}

where $s_{0}$ signals the onset of the continuum, $f_{\pi} \simeq 93$ MeV,
and
\begin{equation}
\rho (s, s', Q^{2})|_{QCD} = \frac{3}{16 \pi^{2}} \;
\frac{Q^{4}}{\lambda^{7/2}} \left[ 3 \lambda (x + Q^{2})
(x + 2Q^{2}) - \lambda^{2} - 5 Q^{2} (x + Q^{2})^{3} \right] \,,
\end{equation}

to one-loop order (and in the chiral-limit), with
\begin{equation}
\lambda = y^{2} + Q^{2} (2x +Q^{2}) \,,
\end{equation}

and $x = s + s'$, $y = s - s'$. Since we are interested in writing
the lowest moment FESR for $F_{\pi}$, i.e.
\begin{equation}
F_{\pi}(Q^{2}) = \frac{1}{f_{\pi}^{2}} \; \int_{0}^{s_{0}} \; dx \;
\int_{-x}^{x} \; dy \; \rho(x,y,Q^{2})|_{QCD} \,,
\end{equation}

rather than a Laplace transform QCD sum rule, the non-perturbative
power corrections entering the OPE are of no concern to us here
(they contribute to higher moment FESR). The integration region in (6) has
been chosen to be a triangle in the (s,s') plane, with base and height
equal to $s_{0}$. Other choices of the integration region, e.g. a
square region of side $s_{1} \simeq s_{0}/\sqrt{2}$, give similar results.
The solution to the FESR (6) is
\begin{equation}
F_{\pi}(Q^{2}) = \frac{1}{16 \pi^{2}f_{\pi}^{2}}
		     \frac{s_{0}}{(1 + Q^{2}/2 s_{0})^{2}} \,.
\end{equation}

Although not evident from (7),
it is important to realize that this analysis is only valid in the region
$Q^{2} \geq 1$ GeV$^{2}$, where one expects a reasonable convergence of
the OPE. This limitation
is of no relevance if one is only interested in the thermal behaviour of
the ratio $F_{\pi}(Q^{2}, T) / F_{\pi}(Q^{2}, 0)$. In any case, as shown
in \cite{fpi0}, Eq.(7) provides a reasonable fit to the experimental data
in the region $Q^{2} \simeq 1 - 4 $ GeV$^{2}$, if $s_{0} \simeq$ 1 GeV$
^{2}$.\\

We now proceed to derive the thermal corrections to (6). The vacuum average
in Eq.(1) is to be replaced by the Gibbs average
\begin{equation}
<<\mbox{A} \cdot \mbox{B} \cdots >> \equiv \sum_{n} \exp (-E_{n}/T)
<n| \mbox{A} \cdot \mbox{B} \cdots |n>/Tr (\exp (-H/T)) \,,
\end{equation}

where $|n>$ is any complete set of eigenstates of the (QCD) Hamiltonian,
e.g. hadronic states, quark-gluon basis, etc.. If one is to extend
smoothly the $T=0$ QCD sum rule program to finite temperature, then the
natural choice for $|n>$ should be the quark-gluon basis, as first
proposed in \cite{bs}. In fact, assuming the validity of the OPE at
$T \neq 0$ and invoking QCD-hadron duality at $T = 0$, raising the
temperature by an arbitrary small amount would induce very little change
in the hadronic spectrum, and in the expectation values of the QCD
operators in the OPE. Hence, it is reasonable to expect that the
inter-relationship between QCD and hadronic parameters effected by
duality will remain valid. An abrupt dissapearance of this
inter-relationship as soon as $T \neq 0$ appears highly unlikely. As to
the validity of the OPE at finite temperature, one is faced with the
same problem as when $T = 0$, i.e. no rigorous proof can be given since
one is not able to solve QCD analytically and exactly. Instead, other
field theory models which are exactly solvable have been used to argue
for the validity of the OPE at $T = 0$ \cite{ope0}.
This analysis has been extended
recently to $T \neq 0$ \cite{opet}, and it shows the same level of
supportive evidence as in the $T = 0$ case.\\

The QCD program at $T \neq 0$  as outlined above leads to some
interesting results. For instance, considering the two-point function
associated with the axial-vector current, a lowest moment FESR relates
$s_{0}(T)$ to $f_{\pi}(T)$ in such a way that $s_{0}(T)$ vanishes at
some temperature $T_{d}$ \cite{s01}-\cite{s02}.
This may be interpreted as the critical temperature
for deconfinement. In fact, with increasing T, and as resonances begin
to melt, the hadronic spectrum should smooth out and $s_{0}(T)$ should
decrease. Close to $T_{d}$ the spectral function is then described
almost entirely by the quark-gluon degrees of freedom at all energies.
Depending on the expression for $f_{\pi}(T)$ used as input, $T_{d}$ could
be smaller or almost the same as the critical temperature for
chiral-symmetry restoration, $T_{c}$. For instance, using chiral
perturbation theory for $f_{\pi}(T)$ leads to $T_{d} < T_{c}$ \cite{s01},
but this input is only valid at low temperatures. A more refined analysis
gives \cite{s02}
\begin{equation}
\sqrt{\frac{s_{0}(T)}{s_{0}(0)}} \simeq \frac{f_{\pi}(T)}{f_{\pi}(0)} \,,
\end{equation}

and $T_{d}$ only slightly lower than $T_{c}$ (given the accuracy of
the method this difference is of not much significance).\\

We have calculated the spectral function (4) at finite temperature using
the Dolan-Jackiw formalism. After substitution in (16),
the result can be expressed as
\begin{equation}
F_{\pi}(Q^{2},T) = \frac{1}{f_{\pi}^{2}(T)} \;
\int_{0}^{s_{0}(T)} \; dx \; \int_{-x}^{x} \; dy \;
\rho(x,y,Q^{2})|_{QCD} F(x,y, Q^{2}, T) \,,
\end{equation}

with
\begin{equation}
F(x,y,Q^{2},T) = 1 - n_{1} - n_{2} - n_{3} + n_{1} n_{2} + n_{1} n_{3} + n_{2}
n_{3} \,,
\end{equation}

\begin{equation}
n_{1} = n_{2} \equiv n_{F} \left(| \frac{1}{2T} \sqrt{\frac{x+y}{2}}|
\right) \,,
\end{equation}

\begin{equation}
n_{3} \equiv n_{F} \left( | \frac{Q^{2} + (x-y)/2}
					   {2T \sqrt{\frac{x+y}{2}}} | \right) \,,
\end{equation}

and $n_{F}$ is the Fermi thermal factor.
In the equations above we have chosen a frame such that
$p_{\mu} = (\omega,{\bf 0})$, and $p'_{\mu} = (\omega',{\bf p'})$, in
which case
\begin{equation}
\omega = \sqrt{\frac{x+y}{2}} \hspace{1cm} , \hspace{1cm} \omega' = \frac{x +
Q^{2}}
{2 \sqrt{\frac{x+y}{2}}} \,.
\end{equation}

We have explicitly checked that the ratio
\begin{equation}
R(T) \equiv \frac {F_{\pi}(Q^{2},T)}{F_{\pi}(Q^{2},0)}
\end{equation}

is essentially insensitive to other choices of frames. For instance, one
may choose $p_{\mu} = (\omega,{\bf p})$, and $p'_{\mu} = (\omega',{\bf - p})$,
which leads to different arguments in the thermal factors, but roughly the
same ratio $R(T)$.
The temperature
behaviour of the ratio (15) is shown in Fig.1
for $Q^{2} = 1$ GeV$^{2}$ (solid curve), and
$Q^{2} = 3$ GeV$^{2}$ (broken curve), where
we have used (9) together with
the results of \cite{s02} for $f_{\pi}(T)$. This result for $R(T)$ is
in nice agreement with the expectation that as the temperature increases,
$F_{\pi}$ should decrease and eventually vanish at the critical temperature
for deconfinement $T_{d}$.\\

Although the OPE breaks down at small values of $Q^{2}$, one may still
extrapolate the ratio (15) into this region just to study the
qualitative temperature behaviour of the root-mean-square radius ratio
$<r^{2}_{\pi}>_{T}/<r^{2}_{\pi}>_{0}$. Doing this, we find  that
this ratio increases monotonically with  T, doubling
at $T/T_{d} \simeq 0.8$, and diverging at the critical temperature. This
divergence of $<r^{2}_{\pi}>_{T}$ may
be interpreted as a signal for quark deconfinement. In fact, the
behaviour of $<r^{2}_{\pi}>_{T}$ can be traced back to the temperature
behaviour of the asymptotic freedom threshold $s_{0}(T)$. As $s_{0}(T)$
decreases with increasing $T$, a signature of quark deconfinement, the
root-mean-square radius of the pion increases. This result is in
qualitative agreement with that obtained in the framework of the
Nambu-Jona Lasinio model \cite{njl}

\subsection*{Acknowledgements}
The work of (CAD) has been supported in part by the John Simon Guggenheim
Memorial Foundation, and that of (ML)
by FONDECYT 0751/92. This
work has been performed in the framework of the FRD-CONICYT Scientific
Cooperation Program.

\subsection*{Figure Captions}

\begin{description}
\begin{itemize}
\item[Figure 1:] Temperature beahviour of the ratio (15) for $Q^{2} = 1$
		     GeV$^{2}$
		     (solid curve), and $Q^{2} = 3$ GeV$^{2}$ (broken curve).
\end{itemize}
\end{description}
\end{document}